# Load distribution and friction torque in four-point contact slewing bearings considering manufacturing errors and ring flexibility

Iker Heras[1], Josu Aguirrebeitia[1], Mikel Abasolo[1], Ibai Coria[1], Iñigo Escanciano[1]
[1] Department of Mechanical Engineering, University of the Basque Country (UPV/EHU), Bilbao, Spain

## ABSTRACT

This work introduces a methodology for the calculation of the load distribution in four-point contact slewing bearings considering ball preload, manufacturing errors and ring flexibility. The model is built by the formulation and minimization of the potential energy of the bearing. Comparing with the rigid rings assumption, the results show that ring deformations involve lower interferences in idling conditions, and have a great effect in the load distribution, but not under external loads. Additionally, a new approach has been proposed for the calculation of the friction torque, which has lower computational cost in comparison to a previous approach by the authors, so more accurate results can be obtained due to refined calculations with no significant cost increase.

**KEYWORDS**
slewing-bearing, four-point contact, manufacturing errors, ring flexibility

## 1. INTRODUCTION

Slewing bearings are used for orientation purposes in several machines like tower cranes, radio-telescopes or solar trackers, amongst others. Another very common application is the yaw and pitch rotation in Wind Turbine Generators (WTGs); this sector has been experiencing a remarkable growth over recent years [1]. However, the sector requests a deeper knowledge of WTG components to gain expertise in the design process and to consequently obtain more reliable and cost effective WTGs.
Figure 1 shows the external loads acting on the slewing bearing: axial load, radial load and tilting moment. The load distribution problem seeks to find how these applied loads are distributed among the different rolling elements, in order to verify that a particular bearing fulfils the design requirements in terms of load carrying capacity and structural stiffness [2]. Moreover, these loads can be used to further calculate the friction torque, which is useful for the dimensioning of the pitch and yaw bearing actuation systems in early design stages. The first approach to solve the load distribution in four-point contact slewing bearings was proposed by Zupan and Prebil [3]; since then, many other works have dealt with this issue [4–7]. The flexibility of the rings has a large influence in the load distribution of slewing bearings, as demonstrated by Aguirrebeitia et al. [6,8,9] and Olave et al. [5] and therefore must be considered for accurate results.

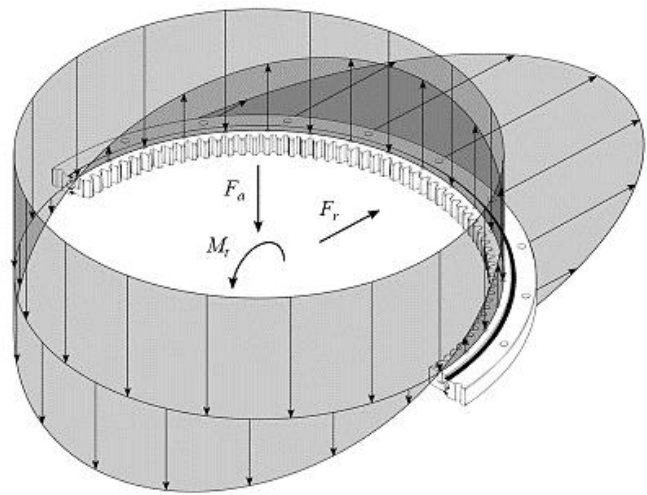

**Figure 1.** Load distribution in slewing bearings under external loads.

To consider the flexibility of the rings, their structural behavior can be simulated by means of stiffness matrices obtained from Finite Element (FE) analysis. These matrices can be obtained through FE static condensation method, also known as superelement technique. A first approach was proposed by Olave et al. [5], who used the mechanism developed by Smolnicki [10] assuming fixed ball-raceway contact points as master nodes. Later, a mechanism more suitable to catch the effect of the contact angle variation was developed by Daidié [11], who linked the center of each raceway to a rigid shell with the dimensions of the contact ellipse by means of rigid beams. In a similar way to Olave et al., Plaza et al. also used the superelement technique to obtain the stiffness matrices of the rings in their study [7], but stablishing their method upon Daidie's mechanism instead of Smolnicki's mechanism, so the stiffness of the rings was reduced to all the nodes of the ball raceway contact surface and not to a single point. The models proposed in these works are established upon the contact geometrical interference problem, while FE analyses are required to simulate rings flexibility.

Besides, recent studies by Starvin and Manisekar [12] and Aithal et al. [13] demonstrated via FE calculations that manufacturing errors can significantly affect the load distribution in large diameter angular contact ball bearings. In a previous manuscript, Heras et al. presented an innovative analytical approach to calculate ball-raceway interferences due to manufacturing errors assuming rigid rings [14]. This model, which will be called BIME (Bearing Interferences due to Manufacturing Errors) in the present work, was later used in combination with FE calculations to evaluate the influence of manufacturing errors and ring stiffness on the idling friction torque value.

Regarding the friction torque calculation in four-point contact slewing bearings, it can be addressed by FE analyses as in [14], but there also exist different analytical approaches [15,16]. The most comprehensive model for this purpose is the one developed by Leblanc and Nélias [17,18], which is based on the work of Jones for angular contact bearings [19] but extended for the four-point contact case. This model was later particularized for the friction torque calculation in slow-speed applications by Joshi et al. [20], being a useful tool for orientation focused bearings. Nevertheless, these formulations assume that full sliding occurs in the ball-raceway contact, which entails some limitations as demonstrated in [21].

The current work proposes a new approach for the load distribution and friction torque calculations in four-point contact slewing bearings. Instead of assuming rigid rings for the load distribution problem as in [14], the new approach considers them flexible by means of stiffness matrices obtained from FE static condensation method; the master nodes are different from those used in previous works [5,7], leading to more accurate and cost-effective results. Having considered the effect of ring flexibility in the load distribution calculation, the friction torque problem can be solved with a rigid-ring FE model, which again will be advantageous with respect to the previous approach [14]. Additionally, the effect of external forces has been included in contrast to previous approach in [14], which was developed for idling condition.

## 2. LOAD DISTRIBUTION MODEL

The following section 2.1 briefly introduces the original BIME model developed for rigid rings in [14], which will be called Rigid-BIME from now on. Then, section 2.2 describes the novel procedure to implement ring flexibility (Flexible-BIME from now on), including the effect of external loads in section 2.3. Finally, section 2.4 presents the validation of new load distribution model.

### 2.1. Model for rigid rings (Rigid-BIME)

The approach in [14] is based on the simulation of the ball-raceway contacts through the mechanism in Figure 2. Depending on the manufacturing errors, a gap (diagonal 1 in Figure 2, i.e. points $P^1$ and $P^3$) or an interference (diagonal 2 in Figure 2, i.e. points $P^2$ and $P^4$) may exist in a contact. The location of the raceway centers ($O^i$) is determined from experimental measurements. These centers are then linked by traction only springs, whose natural length is also a function of the real shape of the raceways. To determine the interferences in the contacts after the assembly of the bearing (after inserting the balls and before applying any external load), the potential energy of the system is formulated, which is a function of the relative position of one ring respect to the other [14]:

$$U_{contact} = \frac{2}{5}\sum_{b=1}^{B}\left[K_{Tot}^{1b}\left(\delta_{Tot}^{1b}\right)^{5/2} + K_{Tot}^{2b}\left(\delta_{Tot}^{2b}\right)^{5/2}\right] \tag{1}$$

Where $B$ is the number of balls, $K_{Tot}^{ib}$ is the total stiffness of the spring $i$ of the ball $b$, and $\delta_{Tot}^{ib}$ is its total elongation (the summation of the interferences in each contact pair). The final relative position will be the one with the minimum associated energy. Therefore, the final spatial configuration, and thus contact interferences, can be determined by minimizing expression (1).

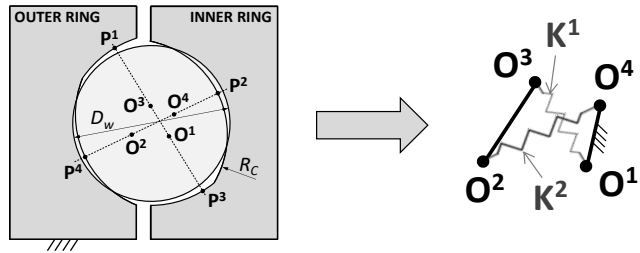

**Figure 2**. Graphical representation of the mechanism of the analytical model for rigid rings case.

### 2.2. Implementation of ring flexibility (Flexible-BIME)

In expression (1) rigid rings are assumed, so only the potential energy associated to ball-raceway contacts is computed. Nevertheless, ring deformations have a great influence when it comes to large diameter bearings such as slewing bearings [5,7,12], so it must be considered to obtain accurate results. For this purpose, the FE static condensation method is used, which condensates the stiffness matrix of a body to a set of preselected Degrees of Freedom (DoF). Selected DoF are those with boundary conditions or loads applied. In order to obtain the stiffness matrix of any ring, a fully parametric FE model was built in ANSYS® (see Figure 3), so bearings with any dimensions could be quickly modelled. Note that the span angle corresponding to each ball will vary depending on the number of balls (see Figures 3b and 3c). This model assumes no gear or holes, since their effect was found to be small. The effect of ring manufacturing errors on the flexibility of the rings is also neglected, given their small magnitude (microns) in comparison with the general dimensions of the ring. The model constraints were defined to represent a real bearing in idling conditions, so that they can deform freely in the space but no rigid body motion is allowed.

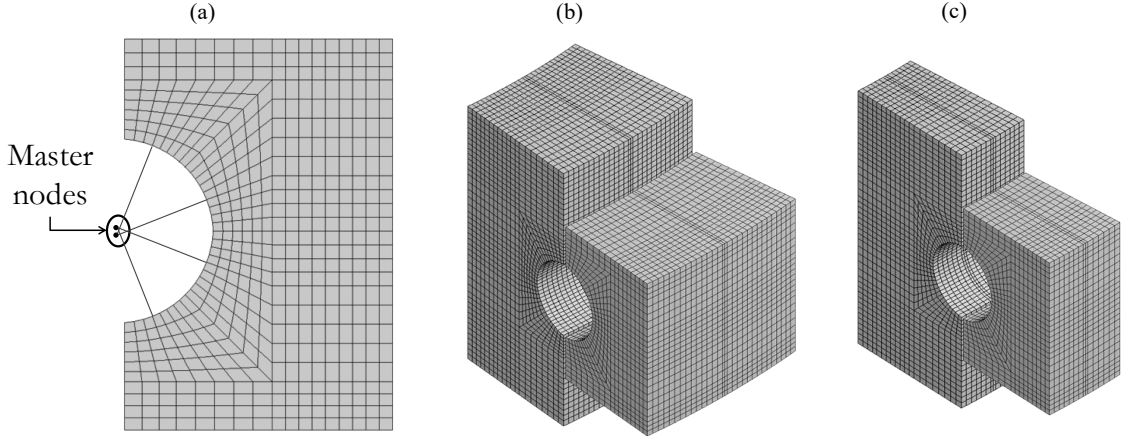

**Figure 3.** FE model: (a) master nodes; (b) sector mesh; (c) sector mesh with lower span angle.

From this FE model, the stiffness matrix of each ring condensed to the centers of the raceways ($O^i$ in Figure 2, master nodes in Figure 3a) can be obtained. Selecting the center of the raceways as master nodes, a more compact stiffness matrix is obtained compared to the one proposed by Plaza [7]. On the other hand, it considers contact angle variation by using Daidié´s mechanism, while the approach by Olave [5], based on Smolnicki's mechanism, had certain limitations: this implies more accurate results when large deformations take place in the mechanism. Furthermore, since no load must be transmitted between the rings in the circumferential direction, only axial and radial degrees of freedom are considered in the master nodes, thus solving the misalignment problems in the mechanism when radial external loads are applied.

Once these stiffness matrices are obtained, they can be implemented in the analytical model. Having selected as master nodes the points $O^1, O^2, O^3$ and $O^4$ in Figure 2, the obtained matrices can be implemented in a straightforward way in the BIME model. Now, the final position of the centers of the raceways will not depend on rigid body displacements of one ring with respect to the other as in previous approach [14], but also on the elastic deformation of the rings. As well as the traction-only springs in Figure 2, rings also store elastic energy when deformed. Consequently, the total potential energy of the system in idling conditions (after assembly, with no external loads applied) will be given by:

$$U_{idling} = U_{contact} + U_{rings} \qquad (2)$$

In order to calculate the elastic energy of the system, the final position of the centers of the raceways is required. When rigid rings were assumed in [14], the coordinates of the points $O^1$ and $O^4$ were fixed. In the new approach the rings are flexible, so these points are no longer fixed. For each ball, the location of these points is obtained adding the elastic deformation to the initial coordinates as follows:

$$\begin{aligned} R_O^i &= R_{Oini}^i + D^{Ri} \\ z_O^i &= z_{Oini}^i + D^{zi} \end{aligned} \quad \text{where } i \in [1,4] \qquad (3)$$

Where $R_{Oini}^i$ and $z_{Oini}^i$ are the initial cylindrical coordinates of the center of the raceway (see Figure 4) corresponding to contact $i$ (see Figure 2), and $D^{Ri}$ and $D^{zi}$ are the radial and axial displacements due to the deformations of the rings in cylindrical coordinates. Same approach must be done to $O^2$ and $O^3$ to consider elastic deformations:

$$\begin{aligned} R_O^i &= R_{Oini}^i(cos\alpha\,sin^2\,\varphi_O^i + cos\beta\,cos^2\,\varphi_O^i) + (X_D + z_{Oini}^i sin\beta)cos\varphi_O^i + \\ &\quad + (Y_D - z_{Oini}^i sin\alpha)sin\varphi_O^i + D^{Ri} \qquad \text{where } i \in [2,3] \quad (4) \\ z_O^i &= z_{Oini}^i cos\alpha cos\beta + R_{Oini}^i(sin\alpha sin\varphi_O^i - sin\beta cos\varphi_O^i) + Z_D + D^{zi} \end{aligned}$$

Where $X_D, Y_D, Z_D, \alpha$ and $\beta$ are the relative rigid body displacements and rotations of the inner ring with respect to the outer ring due to manufacturing errors (see Figure 4). Note that $\varphi$ remains constant for each ball.

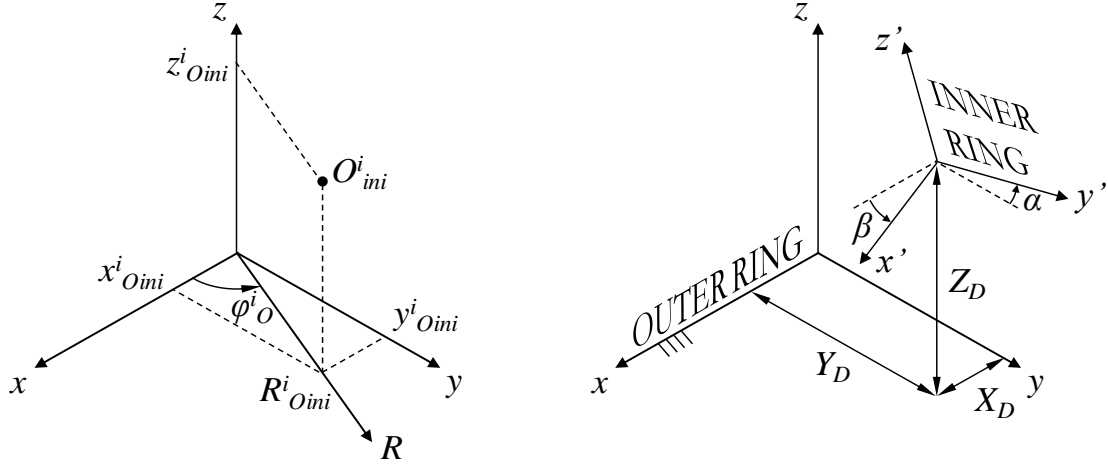

**Figure 4.** Graphical representation of the mechanism of the analytical model with rigid body motion.

In order to calculate the elastic energy stored in the springs that simulate the contact stiffness (Figure 2), their natural length must be formulated first. Since manufacturing errors are being considered, the natural length of the springs may be different from each other. For a given circumferential position, the natural length of spring $i$ is given by:

$$l_N^i = R_C^i + R_C^{i+2} - D_w = R_C^i + R_C^{i+2} - (D_w^{nom} + \delta_P) \quad \text{where } i \in [1,2] \tag{5}$$

Being $D_w$ the ball diameter, equal to the nominal diameter $D_w^{nom}$ plus the preload $\delta_P$, and $R_C^i$ the radius of the raceway. On the other hand, the real length $l$ will be a function of the position of the inner ring:

$$l^i = \sqrt{\left(R_0^i - R_0^{i+2}\right)^2 + \left(z_0^i - z_0^{i+2}\right)^2} \quad \text{where } i \in [1,2] \tag{6}$$

Having both the natural and the real lengths, the summation of the interferences of each contact pair linked by each spring will be calculated according to the following expression:

$$\delta_{Tot}^i = \delta^i + \delta^{i+2} = l^i - l_N^i \quad \text{where } i \in [1,2] \tag{7}$$

Where the contact used in the ball-raceway is hertzian:

$$Q = K\delta^{3/2} \tag{8}$$

Using the approximation developed by Houpert for osculation ratios between 0.89 and 0.99 [22], the next expression is obtained for $E = 2 \cdot 10^5 MPa$ and $\nu = 0.3$:

$$\delta = 5.046 \cdot 10^{-4}(1-s)^{0.2414} \frac{Q^{2/3}}{D_w^{1/3}} \tag{9}$$

Being $E$ in $MPa$, the units to be used in (9) are $[N]$ for $Q$ and $[mm]$ for $\delta$ and $D_w$. The value of $K$ can be formulated from (8):

$$K^i = \begin{cases} \frac{88220\, D_w^{1/2}}{(1-s^i)^{0.3621}} & \text{if } \delta^i > 0 \\ 0 & \text{if } \delta^i \leq 0 \end{cases} \quad \text{where } i \in [1,4] \tag{10}$$

Where $s^i = D_w/(2R_C^i)$ is the osculation ratio of the contact $i$. It is important to point out that, as the springs are traction-only, they do not work for $\delta^i < 0$ case, which represents a gap between the contacting bodies. The total stiffness of spring $i$ that links the raceway centers $i$ and $(i+2)$ is obtained from (8) and (9):

$$\frac{1}{\left(K_{Tot}^i\right)^{2/3}} = \frac{1}{(K^i)^{2/3}} + \frac{1}{(K^{i+2})^{2/3}} \tag{11}$$

The potential energy of the contact ($U_{contact}$) is then calculated according to (1).
On the other hand, the potential energy due to the elastic deformation of the rings ($U_{rings}$) must be obtained in order to calculate the total energy of the system according to equation (2). For such purpose, the FE model in Figure 3 is used to obtain the stiffness matrices of the rings. Considering a bearing with $B$ balls, the dimensions of the stiffness matrix for each ring will be $[4B \times 4B]$. The structure of the matrices is shown below, where $K_{D_2 i_2 b_2}^{D_1 i_1 b_1}$ is the component that relates the degree of freedom $D_1$ ($R$ or $z$) of the raceway center of contact point $i_1$ (1 or 4 for the outer ring and 2 or 3 for the inner ring) and ball $b_1$, with the degree of freedom $D_2$ of the raceway center of contact point $i_2$ and ball $b_2$. Thus, the stiffness matrix for the outer ring is defined as follows:

$$[K_{out}] = \begin{bmatrix} K_{R11}^{R11} & K_{Z11}^{R11} & K_{R41}^{R11} & K_{Z41}^{R11} & & K_{R1B}^{R11} & K_{Z1B}^{R11} & K_{R4B}^{R11} & K_{Z4B}^{R11} \\ K_{R11}^{Z11} & K_{Z11}^{Z11} & K_{R41}^{Z11} & K_{Z41}^{Z11} & & K_{R1B}^{Z11} & K_{Z1B}^{Z11} & K_{R4B}^{Z11} & K_{Z4B}^{Z11} \\ K_{R11}^{R41} & K_{Z11}^{R41} & K_{R41}^{R41} & K_{Z41}^{R41} & \cdots & K_{R1B}^{R41} & K_{Z1B}^{R41} & K_{R4B}^{R41} & K_{Z4B}^{R41} \\ K_{R11}^{Z41} & K_{Z11}^{Z41} & K_{R41}^{Z41} & K_{Z41}^{Z41} & & K_{R1B}^{Z41} & K_{Z1B}^{Z41} & K_{R4B}^{Z41} & K_{Z4B}^{Z41} \\ & & \vdots & & \ddots & & \vdots & & \\ K_{R11}^{R1B} & K_{Z11}^{R1B} & K_{R41}^{R1B} & K_{Z41}^{R1B} & & K_{R1B}^{R1B} & K_{Z1B}^{R1B} & K_{R4B}^{R1B} & K_{Z4B}^{R1B} \\ K_{R11}^{Z1B} & K_{Z11}^{Z1B} & K_{R41}^{Z1B} & K_{Z41}^{Z1B} & & K_{R1B}^{Z1B} & K_{Z1B}^{Z1B} & K_{R4B}^{Z1B} & K_{Z4B}^{Z1B} \\ K_{R11}^{R4B} & K_{Z11}^{R4B} & K_{R41}^{R4B} & K_{Z41}^{R4B} & \cdots & K_{R1B}^{R4B} & K_{Z1B}^{R4B} & K_{R4B}^{R4B} & K_{Z4B}^{R4B} \\ K_{R11}^{Z4B} & K_{Z11}^{Z4B} & K_{R41}^{Z4B} & K_{Z41}^{Z4B} & & K_{R1B}^{Z4B} & K_{Z1B}^{Z4B} & K_{R4B}^{Z4B} & K_{Z4B}^{Z4B} \end{bmatrix} \quad (12)$$

Analogously, the structure of the inner ring stiffness matrix will be:

$$[K_{in}] = \begin{bmatrix} K_{R21}^{R21} & K_{Z21}^{R21} & K_{R31}^{R21} & K_{Z31}^{R21} & & K_{R2B}^{R21} & K_{Z2B}^{R21} & K_{R3B}^{R21} & K_{Z3B}^{R21} \\ K_{R21}^{Z21} & K_{Z21}^{Z21} & K_{R31}^{Z21} & K_{Z31}^{Z21} & & K_{R2B}^{Z21} & K_{Z2B}^{Z21} & K_{R3B}^{Z21} & K_{Z3B}^{Z21} \\ K_{R21}^{R31} & K_{Z21}^{R31} & K_{R31}^{R31} & K_{Z31}^{R31} & \cdots & K_{R2B}^{R31} & K_{Z2B}^{R31} & K_{R3B}^{R31} & K_{Z3B}^{R31} \\ K_{R21}^{Z31} & K_{Z21}^{Z31} & K_{R31}^{Z31} & K_{Z31}^{Z31} & & K_{R2B}^{Z31} & K_{Z2B}^{Z31} & K_{R3B}^{Z31} & K_{Z3B}^{Z31} \\ & & \vdots & & \ddots & & \vdots & & \\ K_{R21}^{R2B} & K_{Z21}^{R2B} & K_{R31}^{R2B} & K_{Z31}^{R2B} & & K_{R2B}^{R2B} & K_{Z2B}^{R2B} & K_{R3B}^{R2B} & K_{Z3B}^{R2B} \\ K_{R21}^{Z2B} & K_{Z21}^{Z2B} & K_{R31}^{Z2B} & K_{Z31}^{Z2B} & & K_{R2B}^{Z2B} & K_{Z2B}^{Z2B} & K_{R3B}^{Z2B} & K_{Z3B}^{Z2B} \\ K_{R21}^{R3B} & K_{Z21}^{R3B} & K_{R31}^{R3B} & K_{Z31}^{R3B} & \cdots & K_{R2B}^{R3B} & K_{Z2B}^{R3B} & K_{R3B}^{R3B} & K_{Z3B}^{R3B} \\ K_{R21}^{Z3B} & K_{Z21}^{Z3B} & K_{R31}^{Z3B} & K_{Z31}^{Z3B} & & K_{R2B}^{Z3B} & K_{Z2B}^{Z3B} & K_{R3B}^{Z3B} & K_{Z3B}^{Z3B} \end{bmatrix} \quad (13)$$

According to matrices (12) and (13), the deformation vectors are defined as follows:

$$\begin{aligned} \{D_{out}\} &= \{D^{R11} \quad D^{Z11} \quad D^{R41} \quad D^{Z41} \quad \cdots \quad D^{R1B} \quad D^{Z1B} \quad D^{R4B} \quad D^{Z4B}\}^T \\ \{D_{in}\} &= \{D^{R21} \quad D^{Z21} \quad D^{R31} \quad D^{Z31} \quad \cdots \quad D^{R2B} \quad D^{Z2B} \quad D^{R3B} \quad D^{Z3B}\}^T \end{aligned} \quad (14)$$

Note that the terms within $\{D_{out}\}$ and $\{D_{in}\}$ are the variables $D^{Ri}$ and $D^{Zi}$ previously defined in equations (3) and (4) to calculate the coordinates of the raceway centers. Thus, the potential energy due to the elastic deformation of the rings can be calculated by means of the following expression:

$$U_{rings} = \tfrac{1}{2}[\{D_{out}\}^T[K_{out}]\{D_{out}\} + \{D_{in}\}^T[K_{in}]\{D_{in}\}] \quad (15)$$

Where $[K_{out}]$ and $[K_{in}]$ are the stiffness matrices of the outer and inner rings respectively, and $\{D_{out}\}$ and $\{D_{in}\}$ are the displacements of the raceway centers due to the elastic deformation of the rings. Consequently, according to equation (2), the total potential energy of the system in idling conditions will be:

$$U_{idling} = \tfrac{2}{5}\sum_{b=1}^{B}\left[K_{Tot}^{1b}\left(\delta_{Tot}^{1b}\right)^{5/2} + K_{Tot}^{2b}\left(\delta_{Tot}^{2b}\right)^{5/2}\right] + \tfrac{1}{2}[\{D_{out}\}^T[K_{out}]\{D_{out}\} + \{D_{in}\}^T[K_{in}]\{D_{in}\}] \quad (16)$$

As it was done for the rigid rings case, the contact interferences can now be obtained by minimizing (16), but the number of unknown variables gets increased from 5 (the 5 rigid body motion degrees of freedom) to 5+8B (being B the ball number) because the terms in equation (14) are unknown. Obviously, this fact increases the computational cost.

One relevant contribution of this paper is that it has been demonstrated that the stiffness matrix of the ring can be obtained from a sector model with the span angle corresponding to one ball, as in Figure 3. Then, the stiffness matrix of the whole ring can be obtained from expanding this matrix, which leads to an equivalent band matrix that provides identical results with a much lower computational cost.

### 2.3. External loads application

The potential energy of the system will now be generalized to account for external loads for any of the two previous models (rigid rings or flexible rings). It is known that the potential energy variation in a system due to an applied conservative load is equal to the negative of the work done by that force. For a load $F$ applied along a displacement $\delta$:

$$F = -\tfrac{dU}{d\delta} \quad \rightarrow \quad dU = -F\,d\delta \quad \rightarrow \quad U = -F\cdot\delta = -W \quad (17)$$

Thus, the total potential energy of the system when external loads are applied can be calculated by deducting the work done by these loads to expression (2):

$$U_{total} = U_{contact} + U_{rings} + U_{loads} = U_{contact} + U_{rings} - W_{loads} \qquad (18)$$

Where $U_{rings} = 0$ for the case of rigid rings. On the other hand, $W_{loads}$ is:

$$W_{loads} = F_a \delta_a + F_r \delta_r + M_t \theta_t \qquad (19)$$

Being $F_a$, $F_r$ and $M_t$ the applied axial and radial loads and the tilting moment respectively, as illustrated in Figure 5. Since the tilting moment $M_t$ is caused by the radial force $F_r$ due to the wind thrust in WTG, they are perpendicular, as illustrated in the figure. According to (18) and (19), the total potential energy of the system when external loads are applied is:

$$U = \frac{2}{5}\sum_{b=1}^{B}\left[K_{Tot}^{1b}\left(\delta_{Tot}^{1b}\right)^{5/2} + K_{Tot}^{2b}\left(\delta_{Tot}^{2b}\right)^{5/2}\right] + \frac{1}{2}[\{D_{out}\}^T[K_{out}]\{D_{out}\} \\ + \{D_{in}\}^T[K_{in}]\{D_{in}\}] - F_a\delta_a - F_r\delta_r - M_t\theta_t \qquad (20)$$

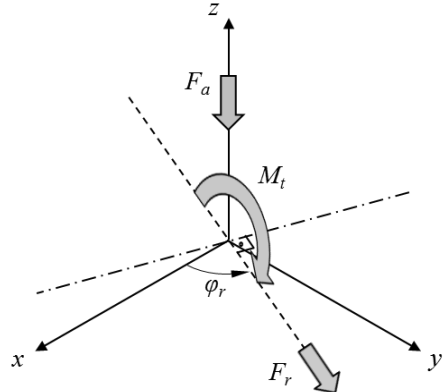

**Figure 5**. Applied external loads.

The external loads in Figure 5 cause $\delta_a$, $\delta_r$, and $\theta_t$ displacements of one ring with respect to the other, taking as the initial position the equilibrium position after the assembly of the bearing. The axial displacement $\delta_a$ is in the $z$ axis, while the radial displacement takes place in the $xy$ plane with a certain $\varphi_r$ angle. The rotation axis for $\theta_t$ is perpendicular to the radial displacement. Of course, the final position of the raceway centers of the mobile ring (the inner one) will be a function of these displacements, taking into account contact interferences and forces, as well as ring deformations. Thus, adding these new variables to the equations (4):

$$R_O^i = R_{Oini}^i\left(\cos\alpha' \sin^2 \varphi_O^i + \cos\beta' \cos^2 \varphi_O^i\right) + \left(X_D + z_{Oini}^i \sin\beta'\right)\cos\varphi_O^i + \\ \left(Y_D - z_{Oini}^i \sin\alpha'\right)\sin\varphi_O^i + \delta_r \cos(\varphi_r - \varphi_O^i) + D^{Ri}$$

$$z_O^i = z_{Oini}^i \cos\alpha' \cos\beta' + R_{Oini}^i\left(\sin\alpha' \sin\varphi_O^i - \sin\beta' \cos\varphi_O^i\right) + Z_D - \delta_a + D^{zi}$$

where $i \in [2,3]$ (21)

Where rotations $\alpha'$ and $\beta'$ will be a function of $\theta_t$:

$$\begin{aligned}\alpha' &= \alpha - \theta_t \sin \varphi_r \\ \beta' &= \beta + \theta_t \cos \varphi_r\end{aligned} \qquad (22)$$

Assuming small displacements, formulas (21) can be simplified as follows:

$$R_O^i = R_{Oini}^i + \left(X_D + z_{Oini}^i \beta'\right)\cos\varphi_O^i + \left(Y_D - z_{Oini}^i \alpha'\right)\sin\varphi_O^i + \\ \delta_r \cos(\varphi_r - \varphi_O^i) + D^{Ri}$$

$$z_O^i = z_{Oini}^i + R_{Oini}^i\left(\alpha' \sin\varphi_O^i - \beta' \cos\varphi_O^i\right) + Z_D - \delta_a + D^{zi}$$

where $i \in [2,3]$ (23)

Once more, minimizing (20), the final position for given external loads (or the reaction forces for certain imposed displacements) could be found. Note that the computational cost will be similar as there are only three more unknown variables than in (16).

## 2.4. Validation of the new load distribution model

This section validates the new approach for the estimation of the ball load distribution considering ring flexibility, i.e. the Flexible-BIME presented in section 2.2, as well as the implementation of external loads in section 2.3. For such purpose, the bearing with the dimensions and axial load carrying capacity in Table 1 has been studied. The dimensions and proportions are similar to those used in bearings for orientation purposes, while the axial load capacity was calculated by means of the analytical model in [6].

| Bearing mean diameter | Ball diameter | Raceway radius | Initial contact angle | Axial static load capacity |
|---|---|---|---|---|
| 1500.00 mm | 35.00 mm | 18.56 mm | 45° | 6318.9 kN |

**Table 1**. Nominal dimensions and axial capacity of the bearing used for validation.

For the validation, the axial stiffness results calculated with the Flexible-BIME model and a full FE model are compared. The FE model is composed by 2 rings as in Figure 3, but modelling the ball with solid elements, with a frictional ball-raceway contact type. The mesh of the FE model is the same as the one used to obtain the ring stiffness matrices in the Flexible-BIME model.

Two completely opposite boundary conditions have been analyzed. In the first case, rigid boundaries are defined, simulating rigid supporting structures. In the second case, frictionless boundaries assume that the rings can deform freely all over the ring-support contact face. As a real working condition will be somewhere between these two extreme cases, the validation of the model under these particular cases will prove that it is suitable for any boundary condition.

Figure 6 shows that the Flexible-BIME and the FE model are in good agreement both with frictionless and rigid boundaries. In this last case, Rigid-BIME results are quite coincident too, because rigid boundary conditions strongly restrict the deformation of the rings in the Flexible-BIME and FE models. As mentioned, realistic boundary conditions would lead to intermediate stiffness values.

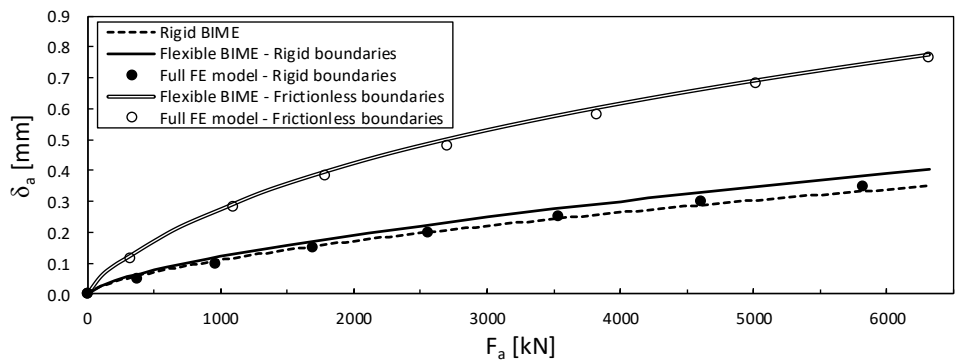

**Figure 6**. Axial stiffness for the bearing in Table 1 according to different models and boundary conditions.

## 3. NEW APPROACH FOR THE FRICTION TORQUE CALCULATION

Recalling the whole procedure, in the previous section the load distribution problem was solved, for which the BIME model is used. The BIME model for rigid rings (Rigid-BIME model explained in section 2.1) is a pure analytical model, whereas the BIME model for flexible rings (Flexible-BIME explained in section 2.2) is a semi-analytical model that combines the analytical model and the FE superelement technique to consider ring flexibility. The results obtained from any of these two models will be implemented in the friction problem, which will provide the contact stress and friction torque results. The friction torque model is a FE model in which the mobile ring is rotated; as a consequence, ball-raceway contact stresses and a corresponding friction torque will be developed. In this friction torque FE model, the rings can be modelled as rigid (Rigid-FEM) or deformable (Flexible-FEM). Whereas in the Rigid-FEM model only ball-raceway deformations are possible (no deformation of the outer faces of the rings is allowed [14]), the Flexible-FEM model has boundary conditions that additionally allow for free ring deformations. A deep discussion on the features of these FE models for friction torque calculation can be found in [14,23]. In summary, and for the sake of clarity, Figure 7 shows a global view of the aforementioned models and simulation alternatives.

In [14], the pure analytical Rigid-BIME approach was used for the estimation of the load distribution; then, the resulting ball-raceway contact forces ($Q$) and angles ($\alpha$) were introduced in the FE rigid model (ID 1 in Table 2). Moreover, the ring flexibility was taken into account when calculating the friction torque by using the Flexible-FEM approach (with flexible rings), which corresponds to ID 2 in Table 2.

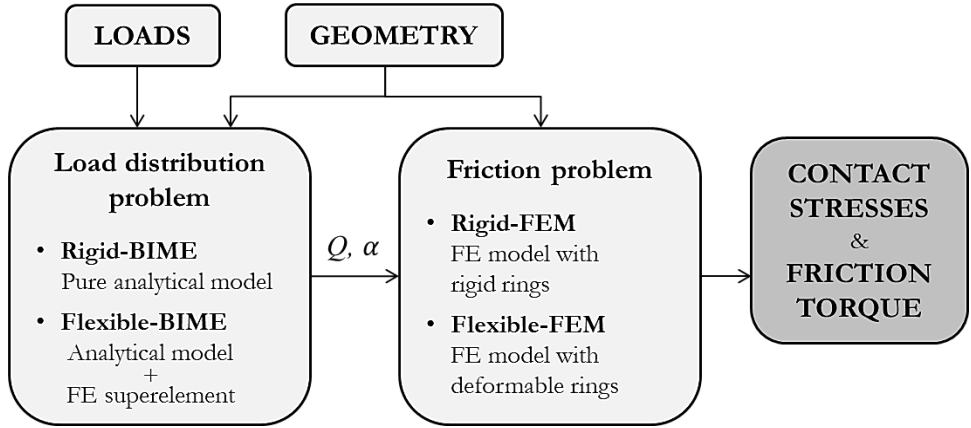

**Figure 7.** Schematic representation of the calculation phases for the developed models.

The present manuscript proposes the alternative ID3 in Table 2, were flexible rings are already considered in the load distribution step by using the Flexible-BIME model, and then Rigid-FEM (with rigid rings) is used to calculate the friction torque.

Thus, while ID1 in [14] did not account for ring flexibility at all, both ID2 in [14] and ID3 in the present work consider this flexibility, but in a different way: ID2 considers it in the friction problem step, and ID3 does it in the load distribution problem step. Even though the results of both ID2 and ID3 approaches are the same as long as the same mesh is used, the new focus used in ID3 offers a number of advantages that will be outlined in the Results and Discussion section.

| ID | Load distribution | Friction problem |
|---|---|---|
| 1 [14] | Rigid-BIME | Rigid-FEM |
| 2 [14] | Rigid-BIME | Flexible-FEM |
| 3 | Flexible-BIME | Rigid-FEM |

**Table 2.** Different approaches for the friction torque calculation.

Before stepping into that next section, a more thorough description of approaches ID2 and ID3 will be presented now, in order to shed light on the particular features of each of them. Regarding the load distribution model, ID2 uses the Rigid-BIME model and ID3 uses the Flexible-BIME one, which implements the ring flexibility obtained via FE superelement technique. In this sense, the load distribution (in terms of ball-raceway interferences, from which the contact forces and angles can be obtained) in ID2 is a consequence of the geometrical interference, which will be larger than the real interference because only ball-raceway deformations are computed. Therefore, when this interference distribution is introduced in the FE model for the friction problem in ID2, the rings of this FE model must be flexible in order to account for ring flexibility (Flexible-FEM model): thus, a first load step in this Flexible-FEM model consists on introducing the interference distribution of the Rigid-BIME model in the Flexible-FEM model, letting it to achieve a new equilibrium state (with the real interference distribution) due to the deformation of the rings. In contrast, this step does not exist in ID3 because the Flexible-BIME model used for the load distribution problem already considers the flexibility of the rings. Obviously, the resulting interference distribution (and therefore the ball contact force and angle distributions) will be the same in both ID2 and ID3 approaches. After that, as mentioned, the mobile ring is rotated with respect to the fixed one, giving rise to contact stresses that will result in a friction torque, the same in both approaches. Once the differences between the two approaches are clear, the results and discussion will show some interesting conclusions and the advantages of the present ID3 approach over the previous ID2 approach.

## 4. RESULTS AND DISCUSSION

### 4.1. Load distribution

Using the measurements of a particular bearing (Table 3) [14], the proposed methodology was applied to calculate ball-raceway interferences in idling conditions assuming both rigid rings (Rigid-BIME, whose results are shown in Figure 8) and considering ring flexibility (Flexible-BIME, whose results are in Figure 9). The comparison reveals the large effect of ring flexibility. Note that contacts 1-3 and 2-4 in Figure 8 to Figure 11 correspond to the contact diagonals illustrated in Figure 2.

| Bearing mean diam. | Ball diameter | Raceway radius | Initial contact angle |
|---|---|---|---|
| 541.00 mm | 25.00 mm | 13.25 mm | 45° |

**Table 3**. Nominal dimensions of the bearing measured in [24].

For the nominal ball (no preload), the average interference decreases from 5μm for rigid rings (Figure 8 top) to 3μm for deformable rings (Figure 9 top). The difference between the maximum and the minimum interferences also decreases from 14μm for rigid rings to 12μm for deformable rings. Thus, ring flexibility leads to lower interferences, and a smoother distribution. This happens because the rings are deformed due to ball-raceway contact loads.

When it comes to ball preload (defined as the increase of its nominal diameter in microns), the effect is more significant. For a preload of +20μm, the average interference decreases from 25μm for rigid rings (Figure 8 down) to 14μm for deformable rings (Figure 9 down). With deformable rings, a given increment in the preload does not involve the same increment in the interferences, as it actually happens with rigid rings, because the higher the preload, the larger the contact loads and therefore the ring deformation is larger. Besides, results show that the interferences distribution gets smoother as the preload increases.

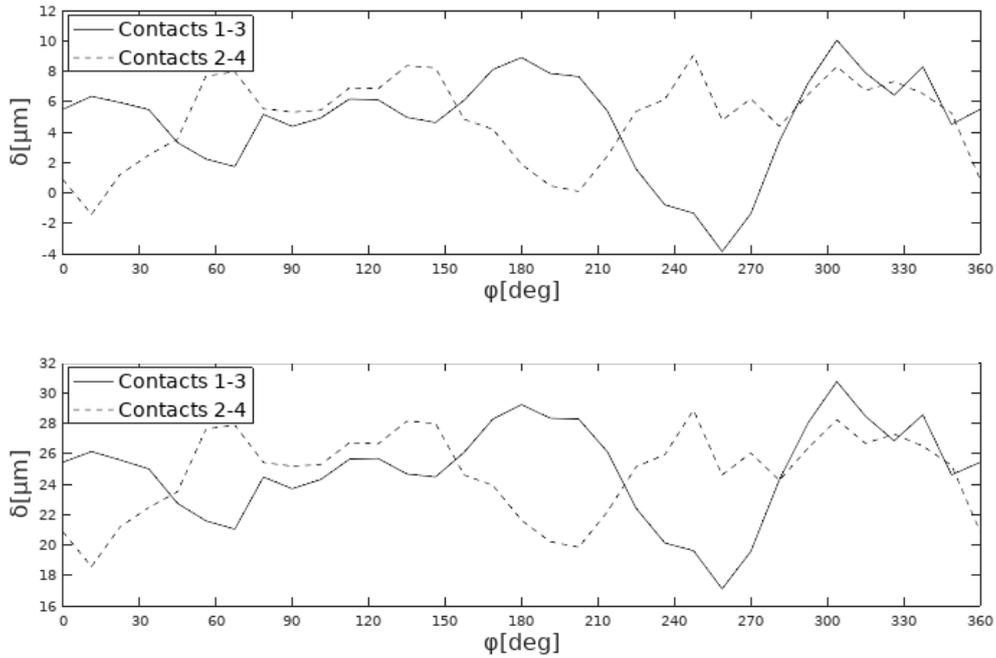

**Figure 8**. Interferences for rigid rings (Rigid-BIME) with 32 balls: nominal ball (top) and +20μm (down).

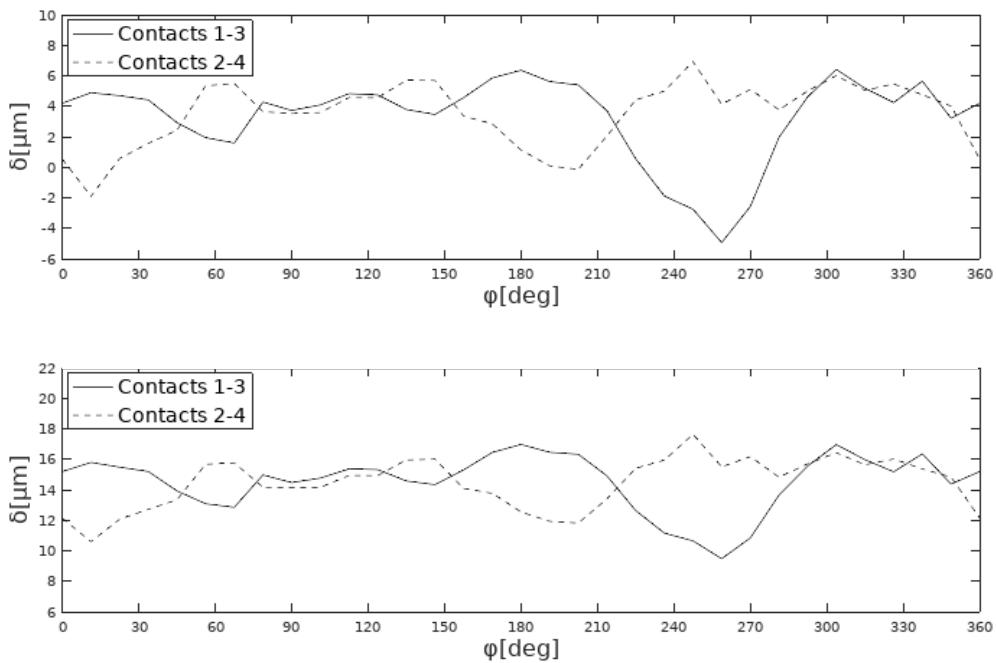

**Figure 9**. Interferences for deformable rings (Flexible-BIME) with 32 balls: nominal ball (top) and $\delta_P = +20$μm (down).

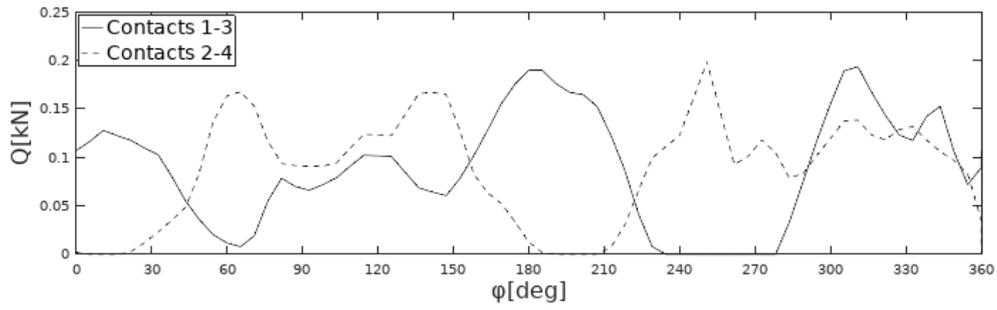

**Figure 10**. Load distribution with no preload and no external loads for 67 balls.

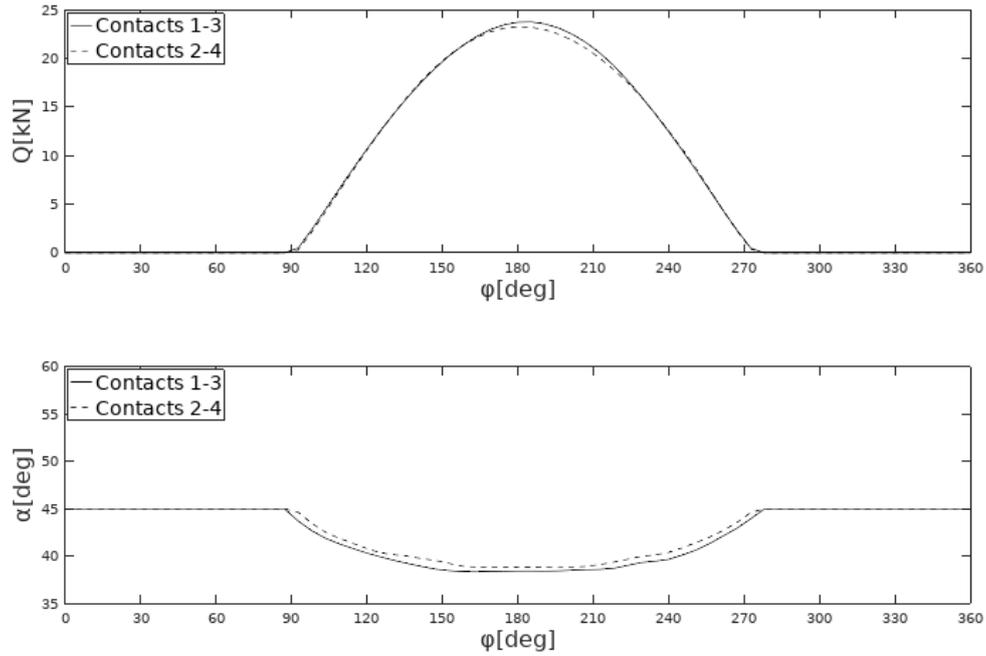

**Figure 11**. Load distribution with no preload and a radial load for 67 balls.

Figure 10 and Figure 11 show the load distribution considering ring flexibility for the idling case and a radial load case, respectively. The radial load is one half the static load capacity of the bearing. From these plots it can be concluded that, under external loads, the effect of manufacturing errors on the load distribution is negligible.

Apart from the fact that the model considers manufacturing errors and ring flexibility, it offers another important advantage. When the stiffness of the ball-raceway contact is simplified by means of a beam-spring mechanism in FE calculations, as done by Smolnicki [10] or Daidié [11], under an external radial load the mechanism leaves this plane, as represented in Figure 12. When this happens, a spurious radial stiffness appears due to the misalignment of the springs. Since the circumferential degree of freedom is not considered in the proposed semi-analytical model, this problem is avoided, thus offering more accurate results for load cases involving radial displacements as in Figure 11. Moreover, once the stiffness matrices have been calculated, the proposed approach is much faster than any FE model even with simplified mechanisms. The developed model only requires FE analysis for the calculation of the stiffness matrices, and then any load case can be quickly solved, while regular FE models require one costly calculation (or load step) for each load case.

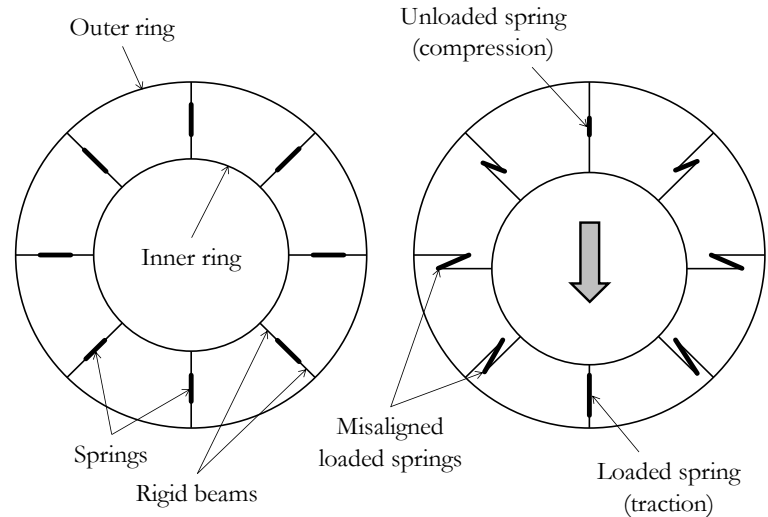

**Figure 12.** Schematic representation of the misalignment of Daidié's mechanism under radial load.

### 4.2. Friction torque

Following the procedure in Figure 7, the results from the load distribution model (ball-raceway contact forces and angles) were used as input data for the friction torque model. The friction torque can be calculated by means of a FE model with rigid (Rigid-FEM) or flexible (Flexible-FEM) rings according to Table 2.

As it has been mentioned, if the same mesh is used in the Flexible-FEM of ID2 approach (Figure 13a) and to obtain the ring stiffness matrix for the Flexible-BIME of ID3 approach (Figure 13b), the same frictional torque value is obtained from both approaches. In this sense, the dark band in Figure 14 shows the idling friction torque results for the studied bearing (Table 3) for different preload levels; the dispersion of the band is due to ball size tolerances. Nevertheless, as ID3 accounts for the ring flexibility in the load distribution problem, while ID2 does it in the friction problem, the computational cost is significantly lower in ID3; this can be verified by comparing ID2 and ID3a (which has the same mesh size as ID2) in Table 4. These times were obtained in a high performance work station, with an Intel® Xeon® E5-2697 v3 @ 2.6GHz processor with 14 physical cores (28 logical) and a RAM of 128GB. Note that the computer time for the contact simulation (FE model) varies depending on the number of active contacts (2 or 4) and the magnitude of the contact deformations. This is why a range is given in the table instead of a fixed value.

Taking advantage of the efficiency of ID3, a finer mesh can be adopted in this approach (Figure 13c instead of Figure 13b), thus obtaining the more accurate results shown in the light band in Figure 14. This mesh refinement involves no significant cost increase as pointed out in Table 4, case ID3b. Note that, as the mesh size decreases, so does the stiffness of the rings and the resulting ball-raceway deformations, which finally leads to a smaller friction torque as Figure 14 proves. The final mesh in Figure 13c has been obtained from a sensitivity analysis.

Another major advantage of approach ID3 over ID2 is that, in contrast to ID2, ID3 already considers the ring flexibility in the load distribution problem. As a consequence, the contact forces and angles obtained are the definite ones, and they could be directly used to solve the friction torque problem with an analytical model similar to the one proposed by Leblanc and Nelias [17,21].

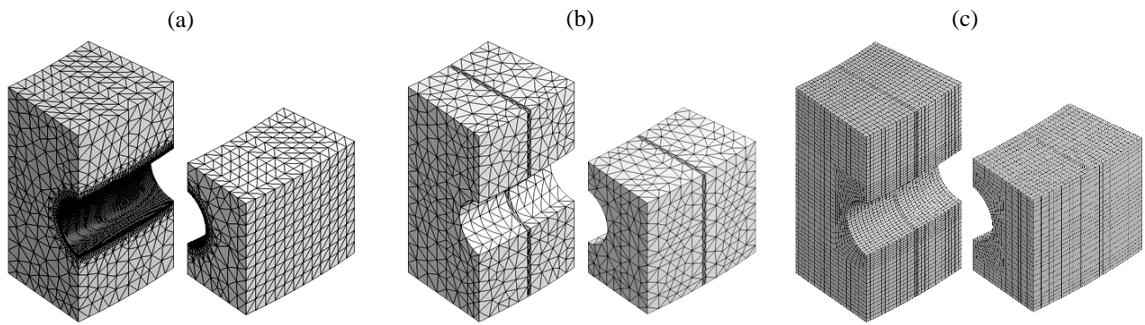

**Figure 13.** Ring mesh: (a) for the friction torque calculation (Flexible-FEM in ID2); (b) for the ring stiffness matrix calculation (Flexible-BIME in ID3) with coarse mesh; (c) for the ring stiffness matrix calculation (Flexible-BIME in ID3) with final mesh.

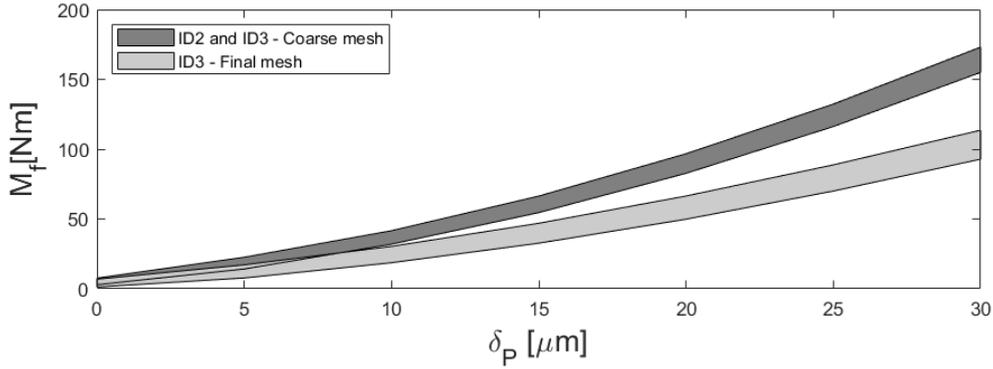

**Figure 14.** Friction torque VS Ball preload for deformable rings: comparison between a coarse mesh and the final mesh.

| ID | Ring stiffness calculation | Load distribution | Contact simulation | Total time |
|---|---|---|---|---|
| 1 | Not required | Rigid-BIME (10 s) | Rigid-FEM (2-8h) | 2-8h |
| 2 | Not required | Rigid-BIME (10 s) | Flexible-FEM (5-20h) | 5-20h |
| 3a | FE-Superelement: Coarse Mesh (10 s) | Flexible-BIME (2 min) | Rigid-FEM (2-8h) | 2-8h |
| 3b | FE-Superelement: Fine Mesh (5 min) | Flexible-BIME (2 min) | Rigid-FEM (2-8h) | 2-8h |

**Table 4.** Computational cost for different calculation cases.

## 5. CONCLUSIONS

This work presents a semi-analytical model to calculate the load distribution in four-point contact slewing bearings considering ball preload, manufacturing errors and ring flexibility both in idling conditions and under axial, radial and tilting moment external loads. Comparing with the rigid rings assumption, the results show that ring deformations involve lower and smoother ball-raceway interferences in idling conditions. The effect of manufacturing errors on the load distribution under external loads is proved to be negligible.

Using this new ball load distribution model, a new approach has been proposed for the calculation of the friction torque. Compared with previous developments by the authors, the new approach obtains the same torque results with a much lower computational cost; in this sense, more refined calculations can be afforded, leading to more accurate results with no significant increase in the cost. Thus, the new approach is an efficient tool for the estimation of load carrying capacity, structural stiffness and friction torque of four-contact point slewing bearings in early WTG design stages.

Finally, the results of this approach can be directly used to feed an analytical model for the friction torque calculation. The authors are currently dealing with this issue, which will give rise to an even more efficient methodology for the estimation of friction torque in four-point contact slewing bearings.

**Acknowledgments**
The authors want to thank Eneko Goikolea for his valuable contributions to this study.
This paper is a result of the close collaboration that the authors maintain with the Basque Bearing Manufacturer Iraundi S.A. The authors also want to acknowledge the financial support of the Spanish Ministry of Economy and Competitiveness through grant number DPI2017-85487-R (AEI/FEDER,UE) and the Basque Government through project number IT947-16.

# LIST OF SYMBOLS

| Symbol | Description |
|---|---|
| $B$ | Number of balls |
| $D_w$ | Ball diameter (including preload) |
| $D_w^{nom}$ | Nominal diameter of the ball (without preload) |
| $D^{Rib}, D^{Zib}$ | Radial ($R$) and axial ($z$) displacements of $O^i$ for the ball $b$ due to the elastic deformations of the rings |
| $\{D_{out}\}, \{D_{in}\}$ | Vectors of the displacements of the raceway centers due to the elastic deformation of the rings |
| $E$ | Young's modulus |
| $F_a, F_r$ | Applied axial ($a$) and radial ($r$) loads |
| $K^i$ | Stiffness of the contact $i$ |
| $K_{Tot}^{ib}$ | Total stiffness of the spring $i$ (which links points $O^i$ and $O^{i+2}$) of the ball $b$ |
| $K_{D_2 i_2 b_2}^{D_1 i_1 b_1}$ | Component of the stiffness matrix of a ring relating the degree of freedom $D_1$ ($R$ or $z$) of the raceway center of contact point $i_1$ (1 or 4 for the outer ring and 2 or 3 for the inner ring) and ball $b_1$, with the degree of freedom $D_2$ of the raceway center of contact point $i_2$ and ball $b_2$ |
| $[K_{out}], [K_{in}]$ | Stiffness matrices of the outer ($out$) and inner ($in$) rings |
| $l^i$ | Final length of the spring $i$ (which links points $O^i$ and $O^{i+2}$) |
| $l_N^i$ | Natural length of the spring $i$ (which links points $O^i$ and $O^{i+2}$) |
| $M_t$ | Applied tilting moment |
| $O^i$ | Final location of the raceway center of the contact point $i$ |
| $O_{ini}^i$ | Initial location of the raceway center of the contact point $i$ |
| $Q^i$ | Normal force in the contact point $i$ |
| $R_C^i$ | Radius of the raceway of the contact point $i$ |
| $(R_O^i, z_O^i, \varphi_O^i)$ | Final position of $O^i$ in the cylindrical coordinate system |
| $(R_{Oini}^i, z_{Oini}^i, \varphi_O^i)$ | Initial position of $O^i$ in the cylindrical coordinate system |
| $s^i$ | Osculation ratio of the contact point $i$ |
| $U_{contact}$ | Potential energy associated to contact deformations |
| $U_{idling}$ | Total potential energy of the system in idling conditions |
| $U_{rings}$ | Potential energy associated to ring deformations |
| $U_{total}$ | Total potential energy of the system considering applied external loads |
| $W_{loads}$ | Work done by the external applied loads |
| $X_D, Y_D, Z_D$ | Rigid body displacements of the inner ring with respect to the outer ring due to manufacturing errors on the $x$, $y$ and $z$ axes |
| $\alpha, \beta$ | Rigid body rotations around $x$ and $y$ axes of the inner ring with respect to the outer ring due to manufacturing errors |
| $\alpha', \beta'$ | Total rigid body rotations around $x$ and $y$ axes of the inner ring with respect to the outer ring due to manufacturing errors and applied external loads |
| $\delta^i$ | Deformation of the contact $i$ |
| $\delta_{Tot}^{ib}$ | Total elongation of the spring $i$ (which links points $O^i$ and $O^{i+2}$) and ball $b$ |
| $\delta_a, \delta_r$ | Axial ($a$) and radial ($r$) rigid body displacements of the inner ring with respect to the outer ring due to applied external loads |
| $\delta_P$ | Ball preload |
| $\theta_t$ | Tilting rotation of the inner ring with respect to the outer ring due to applied external loads |
| $\nu$ | Poisson's ratio |
| $\varphi_r$ | Polar angle that defines the direction of the applied radial force ($F_r$) and tilting moment ($M_t$) |

# REFERENCES

[1] I. Pineda, P. Tardieu, WindEurope, Annual combined onshore and offshore wind energy statistics, 2017, WindEurope. (2018). doi:10.1016/j.preghy.2016.08.046.

[2] ISO 76:2006. Rolling bearings - Static load ratings, 2006.

[3] S. Zupan, I. Prebil, Carrying angle and carrying capacity of a large single row ball bearing as a function of geometry parameters of the rolling contact and the supporting structure stiffness, Mech. Mach. Theory. 36 (2001) 1087–1103. doi:10.1016/S0094-114X(01)00044-1.

[4] J.I. Amasorrain, X. Sagartzazu, J. Damián, Load distribution in a four contact-point slewing bearing, Mech. Mach. Theory. 38 (2003) 479–496. doi:10.1016/S0094-114X(03)00003-X.

[5] M. Olave, X. Sagartzazu, J. Damian, A. Serna, Design of Four Contact-Point Slewing Bearing With a New Load Distribution Procedure to Account for Structural Stiffness, J. Mech. Des. 132 (2010) 21006. doi:10.1115/1.4000834.

[6] J. Aguirrebeitia, J. Plaza, M. Abasolo, J. Vallejo, Effect of the preload in the general static load-carrying capacity of four-contact-point slewing bearings for wind turbine generators: theoretical model and finite element calculations, Wind Energy. 17 (2014) 1605–1621. doi:10.1002/we.1656.

[7] J. Plaza, M. Abasolo, I. Coria, J. Aguirrebeitia, I. Fernández de Bustos, A new finite element approach for the analysis of slewing bearings in wind turbine generators using superelement techniques, Meccanica. 50 (2015) 1623–1633. doi:10.1007/s11012-015-0110-7.

[8] J. Aguirrebeitia, J. Plaza, M. Abasolo, J. Vallejo, General static load-carrying capacity of four-contact-point slewing bearings for wind turbine generator actuation systems, Wind Energy. 16 (2013) 759–774. doi:10.1002/we.1530.

[9] J. Aguirrebeitia, M. Abasolo, R. Avilés, I. Fernández de Bustos, General static load-carrying capacity for the design and selection of four contact point slewing bearings: Finite element calculations and theoretical model validation, Finite Elem. Anal. Des. 55 (2012) 23–30. doi:10.1016/j.finel.2012.02.002.

[10] T. Smolnicki, E. Rusiński, Superelement-Based Modeling of Load Distribution in Large-Size Slewing Bearings, J. Mech. Des. (2007). doi:10.1115/1.2437784.

[11] A. Daidié, Z. Chaib, A. Ghosn, 3D Simplified Finite Elements Analysis of Load and Contact Angle in a Slewing Ball Bearing, J. Mech. Des. 130 (2008) 82601. doi:10.1115/1.2918915.

[12] M.S. Starvin, K. Manisekar, The effect of manufacturing tolerances on the load carrying capacity of large diameter bearings, Sadhana. 40 (2015) 1899–1911.

[13] S. Aithal, N. Siva Prasad, M.S. Shunmugam, P. Chellapandi, Effect of manufacturing errors on load distribution in large diameter slewing bearings of fast breeder reactor rotatable plugs, Proc. Inst. Mech. Eng. Part C J. Mech. Eng. Sci. 0 (2015) 1–12. doi:10.1177/0954406215579947.

[14] I. Heras, J. Aguirrebeitia, M. Abasolo, Friction torque in four contact point slewing bearings: Effect of manufacturing errors and ring stiffness, Mech. Mach. Theory. (2017). doi:10.1016/j.mechmachtheory.2017.02.009.

[15] T.A. Harris, J.H. Rumbarger, C.P. Butterfield, Wind Turbine Design Guideline DG03: Yaw and Pitch Rolling Bearing Life. NREL/TP-500-42362, 2009.

[16] M. Stammler, F. Schwack, N. Bader, A. Reuter, G. Poll, Friction torque of wind-turbine pitch bearings – comparison of experimental results with available models, Wind Energy Sci. Discuss. 2017 (2017) 1–16. doi:10.5194/wes-2017-20.

[17] A. Leblanc, D. Nelias, Ball Motion and Sliding Friction in a Four-Contact-Point Ball Bearing, J. Tribol. 129 (2007) 801–808. doi:10.1115/1.2768079.

[18] A. Leblanc, D. Nelias, Analysis of Ball Bearings with 2, 3 or 4 Contact Points, Tribol. Trans. 51 (2008) 372–380. doi:10.1080/10402000801888887.

[19] A.B. Jones, Ball Motion and Sliding Friction in Ball Bearings, J. Basic Eng. 81 (1959) 1–12.

[20] A. Joshi, B. Kachhia, H. Kikkari, M. Sridhar, D. Nelias, Running Torque of Slow Speed Two-Point and Four-Point Contact Bearings, Lubricants. (2015). doi:10.3390/lubricants3020181.

[21]  I. Heras, J. Aguirrebeitia, M. Abasolo, J. Plaza, Friction torque in four-point contact slewing bearings: Applicability and limitations of current analytical formulations, Tribol. Int. 115 (2017) 59–69. doi:10.1016/j.triboint.2017.05.011.

[22]  L. Houpert, An Engineering Approach to Hertzian Contact Elasticity - Part I, Trans. ASME. 123 (2001). doi:10.1115/1.1308043.

[23]  J. Aguirrebeitia, M. Abasolo, J. Plaza, I. Heras, FEM model for friction moment calculations in ball-raceway contacts for applications in four contact point slewing bearings, in: 14th World Congr. Mech. Mach. Sci. 25-30 Oct., Taipei, Taiwan, 2015. doi:10.6567/IFToMM.14TH.WC.OS18.018.

[24]  I. Heras, J. Aguirrebeitia, M. Abasolo, Calculation of the ball raceway interferences due to manufacturing errors and their influence on the friction moment in four-contact-point slewing bearings, 2017. doi:10.1007/978-3-319-44156-6_1.